\documentclass[preprint,5p,times]{elsarticle}

\usepackage[numbers]{natbib}     
\usepackage{lineno}    
\usepackage{graphicx}
\usepackage{subcaption}
\usepackage[percent]{overpic} 
\usepackage{xcolor}
\usepackage{placeins}
\modulolinenumbers[5]

\begin{document}
\makeatletter
\def\ps@pprintTitle{%
 \let\@oddhead\@empty
 \let\@evenhead\@empty
 \let\@oddfoot\@empty
 \let\@evenfoot\@empty
}
\makeatother
\begin{frontmatter}

\title{Early Prediction of Current Quench Events in the ADITYA Tokamak using Transformer based Data Driven Models}

\author[inst1,inst2]{Jyoti Agarwal}
\ead{202021010@dau.ac.in, jagarwal@ipr.res.in}

\author[inst1]{Bhaskar Chaudhury}
\ead{ bhaskar_chaudhury@dau.ac.in}
\author[inst1]{Jaykumar Navadiya}
\author[inst2]{Shrichand Jakhar}
\author[inst2]{Manika Sharma}

\address[inst1]{Group in Computational Science and HPC, Dhirubhai Ambani University (formerly DA-IICT), Gandhinagar, India - 382007}
\address[inst2]{Institute for Plasma Researach, Gandhinagar, India - 382428}

\begin{abstract}
 Disruptions in tokamak plasmas, marked by sudden thermal and current quenches, pose serious threats to plasma-facing components and system integrity. Accurate early prediction, with sufficient lead time before disruption onset, is vital to enable effective mitigation strategies. This study presents a novel data-driven approach for predicting early current quench, a key precursor to disruptions, using transformer-based deep learning models, applied to ADITYA tokamak diagnostic data. Using multivariate time series data, the transformer model outperforms LSTM baselines across various data distributions and prediction thresholds. The transformer model achieves better recall, maintaining values above 0.9 even up to a prediction threshold of 8-10 ms, significantly outperforming LSTM in this critical metric. The proposed approach remains robust up to an 8 ms lead time, offering practical feasibility for disruption mitigation in ADITYA tokamak. In addition, a comprehensive data diversity analysis and bias sensitivity study underscore the generalization of the model. This work marks the first application of transformer architectures to ADITYA tokamak data for early current-quench prediction, establishing a promising foundation for real time disruption avoidance in short-pulse tokamaks.  
\end{abstract}

\end{frontmatter}

\section{\label{sec:level1}Introduction}

  Tokamak plasmas have demonstrated the ability to sustain themselves over extended periods, when operated within specific limits \cite{artsimovich1972tokamak,paruchuri2025density}. However, exceeding these operational boundaries leads to the rapid onset of major plasma instabilities, culminating in the abrupt termination of the discharge \cite{zakharov2012understanding,boozer2012disruptions,hassanein2010impact}. This phenomenon, known as a plasma disruption in tokamaks, refers to a sudden and unintended loss of plasma confinement, characterized by rapid termination of the plasma current and collapse of magnetic equilibrium. This event leads to the abrupt release of the stored thermal and magnetic energy of plasma, potentially causing severe mechanical and thermal loads on the structural components of the tokamak. Disruptions occur within the millisecond timescale and represent one of the most significant challenges in achieving fusion energy production \cite{yoshino2001characterization}. 
  
  Disruptions are not only a concern due to the loss of plasma energy but also because of the severe damage they can inflict on the plasma-facing components of a tokamak and huge magnetic forces \cite{schuller1995disruptions, boozer2012disruptions, buzio2006determination}. Key plasma parameters such as density, current, and pressure are naturally limited by magnetohydrodynamic (MHD) instabilities. Any attempt to push these parameters beyond their stable operational limits triggers disruptions \cite{nedospasov2008quench}. This process typically begins with the growth of MHD modes that destroy the magnetic flux surfaces, rendering the magnetic field stochastic. As a result, the plasma loses its ability to confine energy, leading to a phase called the thermal quench. During thermal quenching, the plasma energy rapidly dissipates, followed by an increase in its resistivity. This phase is then succeeded by the Ohmic dissipation of the plasma current, culminating in a sudden and dramatic drop in the plasma current, which is called current quench.\\
In the early days of tokamak development, disruptions were seen as manageable challenges due to lower energy levels and forces in smaller devices. However, disruptions are now recognized as one of the most serious challenges in tokamak operation. Beyond immediate energy losses and physical damage, disruptions can lead to long-term operational setbacks, affecting machine availability and increasing maintenance costs. Consequently, prediction and mitigation of disruptions have become high-priority objectives in fusion research. Accurate and timely prediction of disruption can provide opportunities to implement mitigation strategies, such as controlled shutdown procedures thereby reducing damage and improving the overall reliability of fusion devices \cite{raman2018injector, finken2008valve, dhyani2014electrode}.\\
Globally, there is a growing emphasis on developing reliable, data-driven methods for prediction of disruption, utilizing both traditional physics-based modeling and Machine Learning (ML) approaches \cite{boozer2021steering, pustovitov2022forces, strait2019prevention, lehnen2015disruptions}. Recent advances in ML and data-driven approaches have provided a new dimension to this field. In recent years, several research teams have explored data-based models, leveraging the extensive operational datasets from various tokamaks to identify and predict disruption scenarios \cite{rea2019predictor, guo2021prediction, zheng2020predictor, bhrugu, vega2020ai, aymerich2022convolutional, aymerich2023comparison, kates2019predicting, churchill2020convolutional, zhu2021adaptive, zhu2021hybrid, croonen2023investigation,priyanka2024review, wang2021control, niharika}. These models analyse multivariate time-series data from tokamak diagnostics to identify patterns or anomalies that precede disruptions. By employing advanced algorithms, like Long Short Term Memory (LSTM), variants of Covolution Neural Networks (CNN), decision tree algorithms and in some cases hybrid approach to capture long range and multi variant dependencies, researchers aim to develop robust systems capable of real-time disruption forecasting. Additionally, many studies rely on physics-based models, such as DECAF \cite{sabbagh2018event, sabbagh2021event}, which involve solving complex partial differential equations to simulate disruption scenarios. While these models provide valuable insights, they are computationally intensive and time-consuming. 

In this context, India’s contributions to disruption prediction efforts include the data from ADITYA tokamak, operated by the Institute for Plasma Research (IPR) in Gandhinagar \cite{tanna2017overview}. ADITYA has been instrumental in advancing the understanding of plasma behaviour and disruptions \cite{dhyani2014thesis,purohit2020quench, chattopadhyay2006instability,chattopadhyay2009instability, tanna2015mitigation,ghosh2021mitigation} in short duration tokamaks. Over the years, ADITYA has generated a substantial data, providing a valuable resource for data-driven studies using advanced Artificial Intelligence (AI) techniques. \\
Initially, Artificial Neural Networks (ANNs) \cite{sengupta2000forecasting} and LSTM models \cite{agarwal2021sequence} have been explored for disruption prediction in ADITYA tokamak. While these efforts demonstrated the potential of data-driven approaches, they were limited by small and biased datasets, which restricted the generalizability and robustness of the models.
While for classification of disruption events, recent studies have utilised larger and more diverse datasets from ADITYA and ADITYA-U \cite{joshi2024assessment, joshi2022design, muruganandham2024ensemble}. These data sets capture variations in plasma behavior, including differences in quench times. Advanced techniques such as Support Vector Machines (SVM), Deep Neural Networks (DNN), LSTM and stacking ensemble techniques have been employed to analyze these data. These studies have focused on classifying disruption scenarios rather than providing early prediction results. \\
When working with ADITYA data, the primary challenge is ADITYA has a small pulse duration of approximately 120 ms for ADITYA and 300ms for ADITYA-U, significantly shorter than the pulse durations of other international tokamaks used in data-driven models (typically pulse time ~few seconds to 1000 s). Also, given the short duration of plasma shots in ADITYA, identifying the appropriate prediction threshold (the time window between the prediction moment and the actual disruption event) becomes crucial. The prediction threshold should be sufficient for effective disruption mitigation. Prediction accuracy typically decreases as the threshold increases, making it challenging to maintain both timeliness and reliability in forecasting. In addition to accuracy which signifies the overall correctness of the predictions, precision (the measure of correctness in predicting disruption scenario) and recall (ability to predict disruption effectively) are two important measures to measure reliability of a model for prediction studies. Recall is a critical metric for disruption prediction in tokamaks due to the severe operational consequences of missed disruptions for big tokamaks \cite{croonen2023investigation,rea2019predictor}. Disruptions can cause significant structural damage, electromagnetic stress and costly downtime, making it essential to minimize false negatives. High recall ensures that the prediction model identifies as many actual disruptions as possible, enabling timely activation of mitigation strategies. Given that disruptions are rare events, recall is especially important in evaluating the ability of a model to detect these critical occurrences. Prioritizing both, recall and precision, ensures the safety and reliability of tokamak operations. Thus, a model with high recall values in addition to good precision will be a good choice for tokamak, which ensures the minimum miss of a disruptive event. Therefore, a comprehensive study is essential to evaluate model performance in terms of accuracy, precision and recall for different threshold values. A relatively stable model with a prediction threshold of up to 10 ms can be a suitable choice for short-duration plasma machines like ADITYA. As for ADITYA, the complete plasma shot lasts for only 120 ms and recent developments in particle injection through electromagnetic means suggest that a prediction threshold of 5–10 ms may be sufficient for effective disruption mitigation \cite{raman2018injector}.\\
In this study, we introduce the transformer encoder model as a novel approach for prediction of early current quench in ADITYA tokamak. The transformer encoder model, a relatively recent development in the field of ML, offers a promising alternative for disruption prediction. Originally designed for natural language processing and speech recognition, transformers have demonstrated superior performance in time-series prediction tasks \cite{zeyer2019comparison}. Their ability to capture long-range dependencies and handle multivariate data makes them well-suited to the complexity of tokamak disruptions, where dependencies among plasma parameters are both intricate and span multiple time scales. Moreover, transformers are generally more stable during training and require fewer iterations than LSTM, making them advantageous in scenarios with limited knowledge of ongoing phenomena inside tokamak plasma which is a common challenge in tokamak research. Earlier work which uses the transformer model for disruption related study is very limited \cite{spangher2025disruptionbench} and it is the first time a transformer is being used for ADITYA tokamak data. This work aims to evaluate the ability of the transformer model to forecast current quench and compare its performance with that of LSTM-based models. The ADITYA data set used in this study is significantly larger and more diverse than those used in previous ADITYA disruption prediction studies, covering a wide range of plasma quench times and disruption causes. 

A detailed explanation of the data, the adopted methodology, data collection and analysis procedures, data labeling, data diversity analysis as well as the model architectures and training process employed in the study are provided in section 2.  In Section 3, the results of the study are presented followed by conclusion in section 4.

\FloatBarrier
\begin{figure*}[!h]
    \centering
    \includegraphics[width=0.8\textwidth, height=9cm]{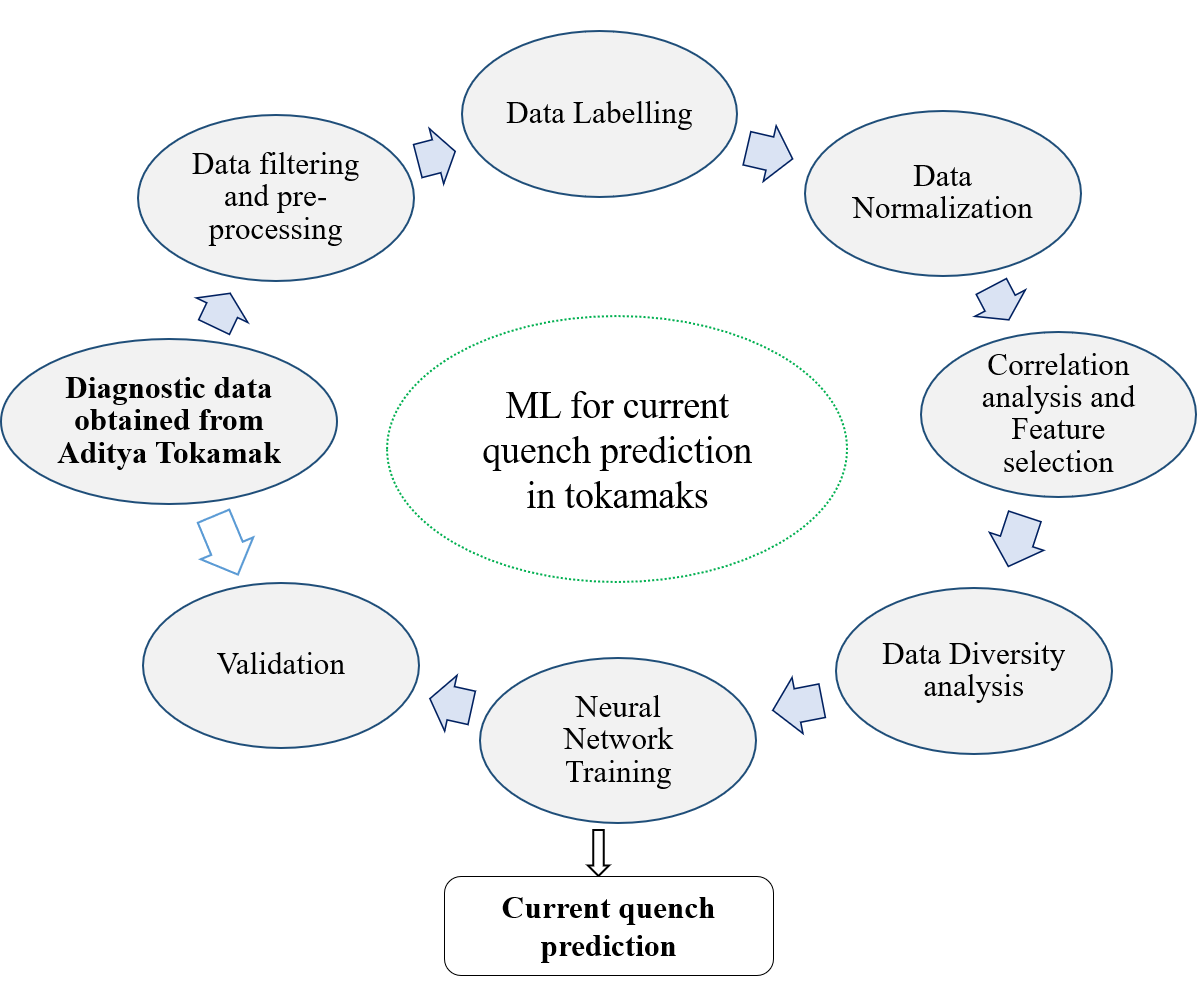}
    \caption{Complete workflow to develop and testing of transformer model for current quench prediction in ADITYA tokamak}
    \label{Figure:methodology}
\end{figure*}
\section{\label{sec:level1}Methodology}

The development of an early current quench predictor for the ADITYA tokamak proposed in this paper involves a systematic workflow.  In the first step, data are collected from various diagnostic channels for a specified range of shots, followed by validation and filtering. The filtered data are then labeled as either disruptive or non-disruptive. The labeled data are then normalized, and key features are selected through correlation analysis. This is followed by an assessment of the diversity of the data, leading to the preparation, training and validation of the predictive model. The entire process is described in figure \ref{Figure:methodology}, with detailed descriptions of each step provided in subsequent subsections.\\

\subsection{\label{sec:level1}ADITYA diagnostic data}

Over three decades of plasma experiments on the ADITYA tokamak, a wealth of data has been generated \cite{bhatt1989aditya, tanna2017overview, raju2000mirnov, atrey2018mhd,bora2002sst}. This dataset includes key diagnostics such as loop voltage (Vloop), plasma current, bolometer readings, C-III and H$\alpha$ radiation, hard and soft X-rays, Mirnov coil outputs, and radiation profiles. These parameters provide a comprehensive understanding of plasma discharge performance, impurity levels (arising from erosion of carbon limiters and other first wall materials) and the intricate dynamics within the tokamak. Visible light spectrometers and mono-chromators are instrumental in monitoring discharge performance and impurity concentrations. Hard X-rays, resulting from collisions of runaway electrons with the first wall or limiters, offered critical diagnostic information. Soft X-rays captured phenomena such as m=1 instability at the q=1 flux surface and tearing instabilities driven by current density gradients, which manifest as filamentation and island formation. The rotation of these islands have been detected using soft X-ray diagnostics. Magnetic field measurements, recorded via Mirnov coils, captured oscillations typically observed under high plasma density or current conditions, providing further insights into plasma behavior.  

\subsubsection{\label{sec:level1}Data collection and analysis}
\begin{figure}[h]
    \centering
    \includegraphics[width=1\linewidth]{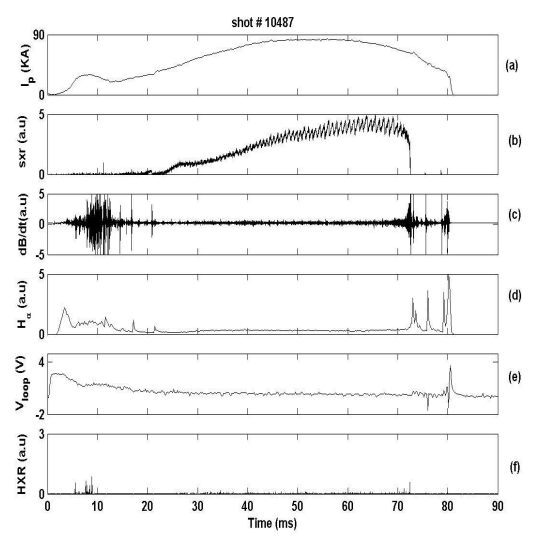}
    \vspace{-5mm}
    \caption{Time evolution of plasma current (a), Soft x-ray signal of central chord (b), Mirnov oscillation (c), H$\alpha$ spectrum (d), loop voltage (e) and hard x-ray spectrum (f) in ADITYA discharge \cite{chattopadhyay2009instability}}
    \label{fig:shot}
\end{figure}

For this study, data is collected from eight diagnostic channels- loop voltage (Vloop), plasma current, bolometer readings, C-III and H$\alpha$ radiation, hard and soft X-rays, and Mirnov coil outputs as shown in Figure \ref{fig:shot}. A detailed correlation analysis has been performed \cite{agarwal2024labelling} and it helps to understand how changes in one variable may be related to changes in another. While correlation analysis helps identify statistical associations between parameters, it does not imply causation. However, it offers a practical starting point for diagnostic feature selection. In this case, correlation analysis established the relation between plasma parameters with plasma current. This analysis was conducted to identify significant plasma parameters that need to be studied extensively to identify disruption precursors.  Correlation between various parameters shows that the outputs of the bolometer and the SXR, HXR and the radiations from H alpha and C-111 are correlated to the plasma current. Based on this result, six diagnostic signals (plasma current, SXR, HXR, Bolo, H alpha, and C-111) are considered here for model input.
\subsubsection{\label{sec:level1}Disruptive and non-disruptive shots}
\begin{figure}[h!]
    \centering
    \includegraphics[width=0.9\linewidth]{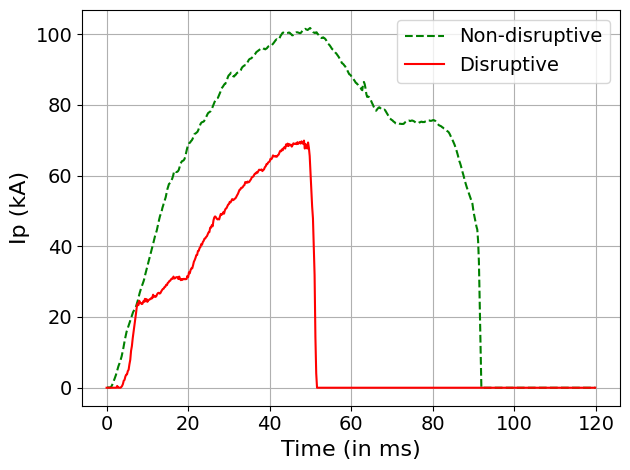}
    \caption{Temporal evolution of plasma current for representative disruptive and non-disruptive discharges in the ADITYA tokamak.}
    \label{fig:dis-nonDis}
\end{figure}

In disruptive cases, current rapidly decays, typically within a few tens of milliseconds, as shown in figure \ref{fig:dis-nonDis} by red line. This decay occurs due to a sharp rise in plasma resistivity following the thermal quench, as the plasma temperature drops and the plasma becomes less conductive. While in non-disruptive cases plasma current does not decay rapidly, instead shows a gradual trend as shown in figure \ref{fig:dis-nonDis} by dashed green line. The overall observation also suggests that there are significant data, in which current quenches before 65 ms and are disruptive in nature.  
\subsubsection{\label{sec:level1}Data labeling}

\begin{figure}[htbp]
    \centering
    \begin{subfigure}{\linewidth}
        \centering
        \includegraphics[width=0.8\linewidth]{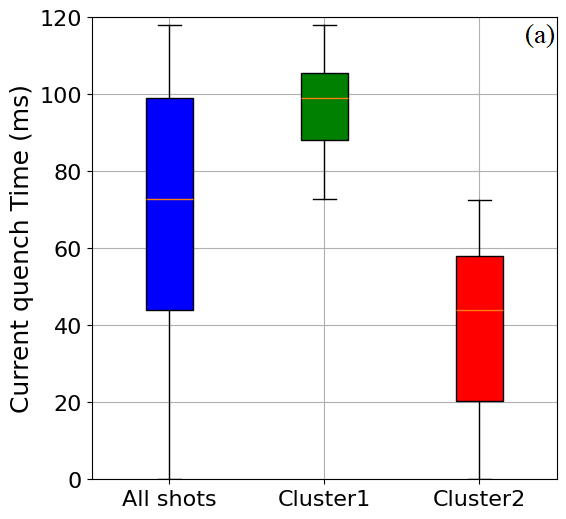} 
        \label{fig:subfig1a}
    \end{subfigure}    
    \vspace{0.1cm}  
    
    \begin{subfigure}{\linewidth}
        \centering
        \includegraphics[width=0.8\linewidth]{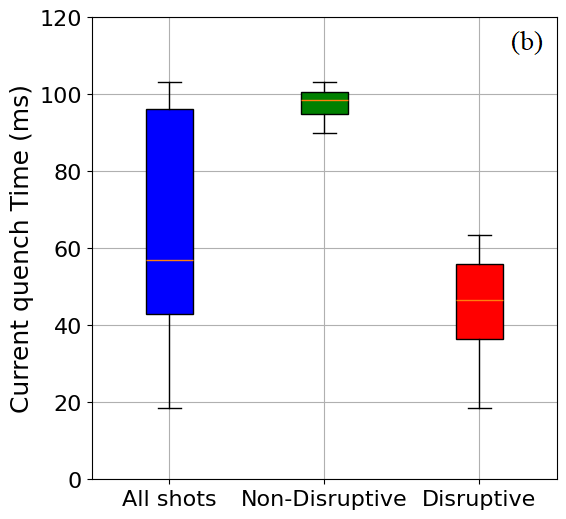}
        \label{fig:subfig1b}
    \end{subfigure}    
    \caption{(a) Distribution of current quench time before outlier removal (b) Distribution of current quench time after outlier removal from experimental data of ADITYA tokamak}
    \label{fig:combined}
\end{figure}

The dataset spans approximately 120 ms of time-series data for selected experimental shots consisting of raw experimental data of 1407 experiments. In our study, data needs to be labelled in two categories, disruptive and non-disruptive. As plasma current quench is an important precursor of disruption, here plasma current quench times are used to label the data. Firstly, two clusters are defined based on the median value of current quench time which is $\approx$72 ms in this case. After dividing the complete dataset into two clusters, outliers are removed in order to create a clear boundary between disruptive and non-disruptive cases.  Outliers are removed by trimming data points that fell outside the ±25$\%$ range of the central value (or median). Following outlier removal, the dataset is reduced to 415 disruptive and 310 non-disruptive shots. The distribution of all shots and the selected shots are shown in Figs. \ref{fig:combined}(a) and (b).

The outlier removal process led to the exclusion of several data points with relatively high quench times. These longer fall-time events typically correspond to non-disruptive terminations or controlled plasma shutdowns. Since most disruptive events exhibit rapid current decay, statistical outlier detection methods, such as interquartile range filtering, tend to classify slow quenches as outliers due to their position in the upper tail of a skewed distribution. This removal is purely mathematical which helps define a sharper boundary between disruptive and non-disruptive events. Notably, this refinement improves the reliability of identifying truly non-disruptive cases and enhances the quality of the dataset, yielding 415 disruptive and 310 non-disruptive cases. The disruptive shots are expected to exhibit more pronounced anomaly signatures, making them more informative for detailed analysis. 

\subsubsection{\label{sec:level1}Data pre-processing}

To ensure uniformity in the time series data being used for this study all the datasets such as SXR, HXR, H$\alpha$ etc. are rescaled to the sampling rate of plasma current, which is 5 kHz sampling rate, preserving essential information resulting in 600 time steps per shot for all the diagnostic signals used for this study. Negative plasma current values are replaced with zero to maintain physical consistency. All parameters are normalized to a range between 0 and 1 to avoid an unnecessary inhomogeneous distribution of the feature weights.
\subsubsection{\label{sec:level1}Data diversity analysis}
Diversity analysis helps design robust algorithms that adapt to a broad spectrum of plasma phenomena, ensuring improved accuracy in prediction models. Furthermore, diversity analysis identifies universal trends and unique behaviours, facilitating cross-device validation and learning. This approach enhances the development of transferable models for disruption prediction across different tokamak devices. Among various methods for quantifying dataset diversity, the Euclidean distance is a simple and effective metric, especially for time-series data. Thus to measure the diversity of data, the Euclidean distance method is used and average Euclidean distance is calculated as per below equations (1) and (2).
\begin{figure}[htbp]
    \centering
    \begin{subfigure}{\linewidth}
        \centering
        \includegraphics[width=0.8\linewidth]{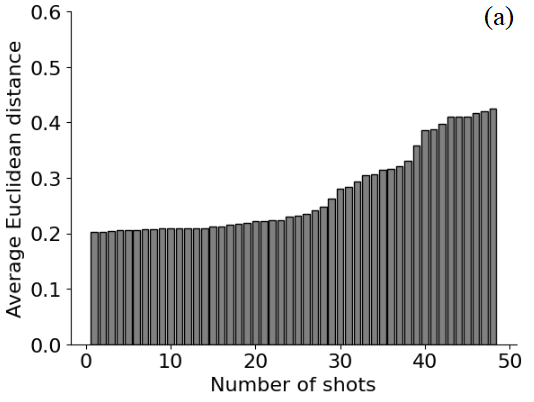} 
        \label{fig:subfig2a}
    \end{subfigure}    
    \vspace{0.1cm}  
    
    \begin{subfigure}{\linewidth}
        \centering
        \includegraphics[width=0.8\linewidth]{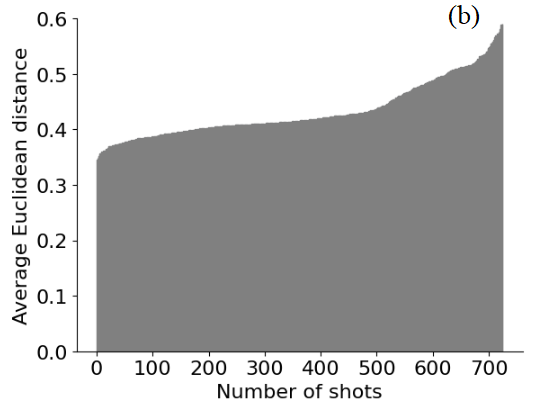}
        \label{fig:subfig2b}
    \end{subfigure}    
    \caption{Disruption studies in ADITYA Tokamak:(a) Data diversity in earlier work \cite{agarwal2021sequence} and (b) Data diversity in current work}
    \label{fig:Eu1}
\end{figure}
\begin{equation}
d_E(\mathbf{Ip}_i, \mathbf{Ip}_j) = \sqrt{ \frac{(\mathbf{Ip}_i - \mathbf{Ip}_j)(\mathbf{Ip}_i - \mathbf{Ip}_j)^T}{N} }
\label{eq:euclidean}
\end{equation}

\begin{equation}
d(\mathbf{Ip}_j) = \frac{ \sum_{{i=1, \\ i \ne j}}^{n} d_E(\mathbf{Ip}_j, \mathbf{Ip}_i) }{(n-1)}
\label{eq:avg_distance}
\end{equation}
Here $Ip_i,Ip_j$ represents the plasma current value vectors of different experimental shots having N points in a single time series and (n) is the total number of experiments being used for study. Euclidean distance in a data frame is a measure of similarity or dissimilarity between two data points. It calculates the straight-line distance between the points, considering all feature values simultaneously. A smaller Euclidean distance indicates that the data points are more similar, while a larger distance suggests greater dissimilarity between the points. 
Diversity for the data collected for the current work is compared with the diversity of data used by A Agarwal et.al. \cite{agarwal2021sequence}, where they have used a total 125 shots from ADITYA out of which 42 shots are used as test shots having 36 disruptive and 6 non-disruptive shots. The values of average Euclidean distance for the data collected for the current work and data segregated as per the description provided in earlier work \cite{agarwal2021sequence} is shown in Figure \ref{fig:Eu1} (a) and (b).

\begin{figure*}[!h]
    \centering
    \includegraphics[width=0.8\textwidth, height=12cm]{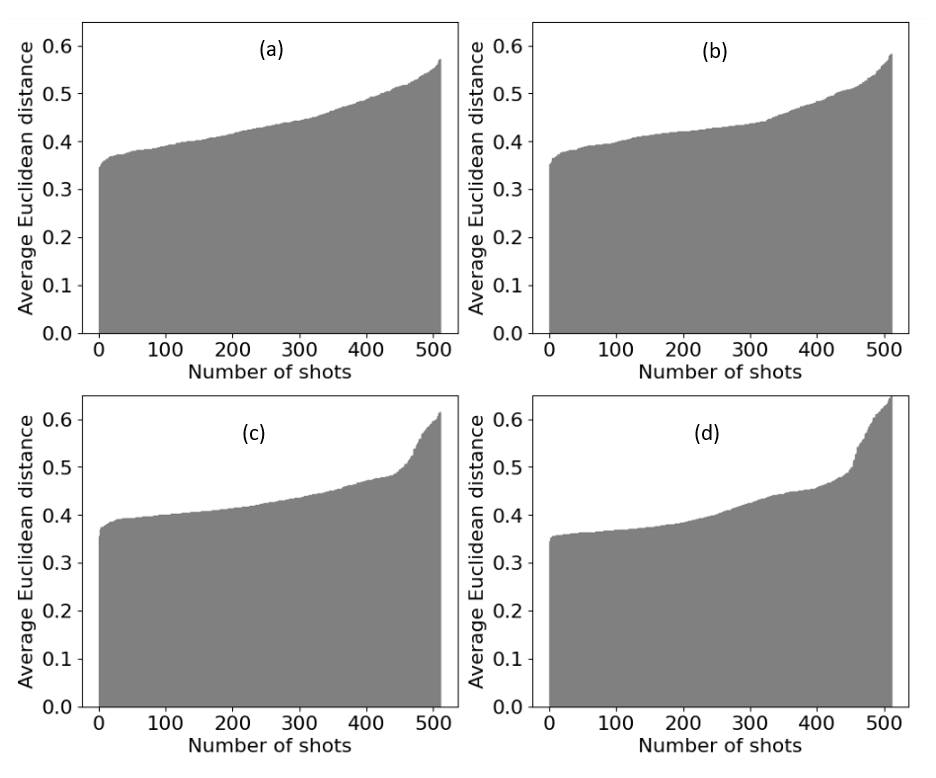}
    \caption{Euclidean distances between plasma current signals in the experimental dataset from ADITYA used for this study, shown for different class distributions: (a) Natural ratio of Disruptive shots and Non-disruptive shots (Unbiased), (b) 60$\%$ Disruptive shots (Mildly biased), (c) 70$\%$ Disruptive shots (Biased) and (d) 80$\%$ Disruptive shots (highly biased)}
    \label{fig:Eu}
\end{figure*}

The analysis reveals that earlier studies focussed on ADITYA plasma disruption prediction relied on low-diversity datasets, which limited their generalization. In contrast, the dataset used in the current study is significantly more diverse. Additionally, further analysis is conducted using datasets with varying levels of bias toward disruptions, which is later employed to evaluate the performance of the transformer model explored in current work. The study with a biased dataset is carried out using a subset of filtered data having a total 510 shots. This choice is made to maintain an adequate number of non-disruptive shots in each scenario while keeping the total dataset size constant, ensuring fairness and uniformity in the evaluation. The results of the diversity analysis are presented in figure \ref{fig:Eu}, in which (a) represents the average Euclidean distance of a dataset having natural ratio of disruptive and non-disruptive shots after filtration of outliers, which is 57$\%$ disruptive and rest non-disruptive shots. Similarly other (b), (c) and (d) are showing the average Euclidean distances of datasets having disruptive shots 60$\%$, 70$\%$ and 80$\%$ respectively.

The study indicates that while the overall distribution of the dataset remains largely unaffected by the degree of bias introduced, the impact of biasing becomes evident when examining the data population in relation to Euclidean distances. Specifically, as the level of bias increases, there is a noticeable rise in the proportion of data points with lower Euclidean distances. This suggests that increasing bias leads to a clustering effect in the feature space. This increased homogeneity within the dataset likely simplifies the patterns that the prediction model needs to learn, thereby may lead to improving the performance of the model. In summary, the study demonstrates that while biasing does not significantly alter the overall data distribution, it increases the density of low-distance data clusters. This, in turn, may help justify the observed performance improvement in predictive models, particularly in tasks that benefit from reduced variability within the training data.

\subsection{\label{sec:level1}Methods, model training and evaluation metrics}
After cleaning and pre-processing the collected data, deep learning models are employed to predict early current quench events. The study utilizes two architectures, LSTM networks and transformer models. Following the development of these models, a validation analysis is conducted using evaluation metrics to assess their performance. Detailed descriptions of the model architectures, training processes, validation methods, and evaluation metrics are provided in the subsequent sections.

\subsubsection{\label{sec:level1}LSTM architecture} 
\begin{figure}[h!]
    \centering
    \includegraphics[width=1\linewidth]{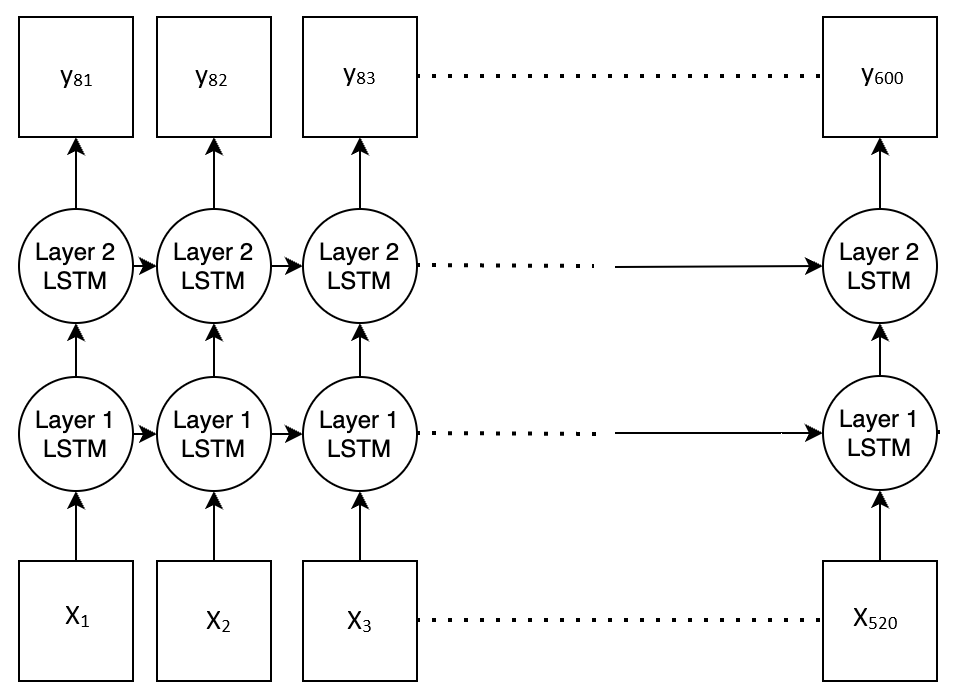}
    \caption{LSTM model architecture for predicting disruptions with 2 layers of LSTM (X is input matrix and y is output vector. $X_1$ is input at time-step 1, $y_{81}$ is output at time-step 81)}
    \label{fig:lstm-architecture}
\end{figure}

To validate the methodology, the LSTM network used in this work is evaluated against previously published results \cite{agarwal2021sequence}. LSTM is a specialized recurrent neural network (RNN) architecture widely used for time-series forecasting tasks. Unlike traditional RNNs, LSTMs are designed to effectively capture long-term dependencies in sequential data. LSTM networks are equipped with memory cells that enable them to selectively retain or forget information over time. This unique capability allows LSTM to learn and preserve essential information across long sequences, addressing the vanishing gradient problem often faced by standard RNNs. 
The architecture of the LSTM model used in this study is illustrated in figure \ref{fig:lstm-architecture}. Diagnostic signals are sampled at 0.2 ms intervals, providing 600 time steps for each 120 ms data segment. The model is trained to predict disruptions identified by current quenching up to 16 ms in advance. To achieve this, inputs $X_t$ from time steps 1 to 520 are aligned with labels $y_{t+80}$ corresponding to time steps 81 to 600. The input $X_t$ is a rectangular matrix comprising six diagnostic signals such as plasma current, soft and hard X-ray outputs, bolometer output, and C111 and H$\alpha$  radiations, serving as the characteristic matrix of the model. 
Where $X_t \in \mathbb{R}^{t \times f}$ , t is the number of time steps and f is the dimensionality of features at each time step, which is 6 for this case.
At each time step $t$, the input sequence is used to predict the output at time step $t+80$ (denoted as $y_{t+80} \in \mathbb{R}^t$). This output, a vector of 80 values , is generated by the LSTM layers as hidden state. This hidden state $y_{t+80}$ is passed through a fully connected layer to produce the final prediction T, likelihood of disruption as shown in equation \ref{T_cal}.

\begin{equation}
T = \sigma(W y_{t+80} + b)
\label{T_cal}
\end{equation}

Here, $W$ is the weight matrix and $b$ is the bias for the output layer, while $\sigma$ denotes the sigmoid activation function. For time series forecasting, the model prediction is based on the input sequence $X_t = \{x_t, x_{t-1}, \dots, x_{t-80+1}\}$,which is compressed by the sigmoid activation function to a single scalar output $T \in [0, 1]$, representing the likelihood of a disruption. This value $T$ is compared with a target value for disruption identification. In this formulation, the target value encodes the proximity to a disruption event, values close to 0 indicate normal plasma operation, while values approaching 1 signify an imminent disruption. A value of 0.5 is selected to trigger an alarm, any predicted value $T > 0.5$ is interpreted as an indication of a possible disruption in the upcoming time window. This value is conservatively chosen to reflect the increased probability of a current quench event.

\subsubsection{\label{sec:level1}Transformer architecture}

\begin{figure}[h!]
    \centering
    \includegraphics[width=1\linewidth]{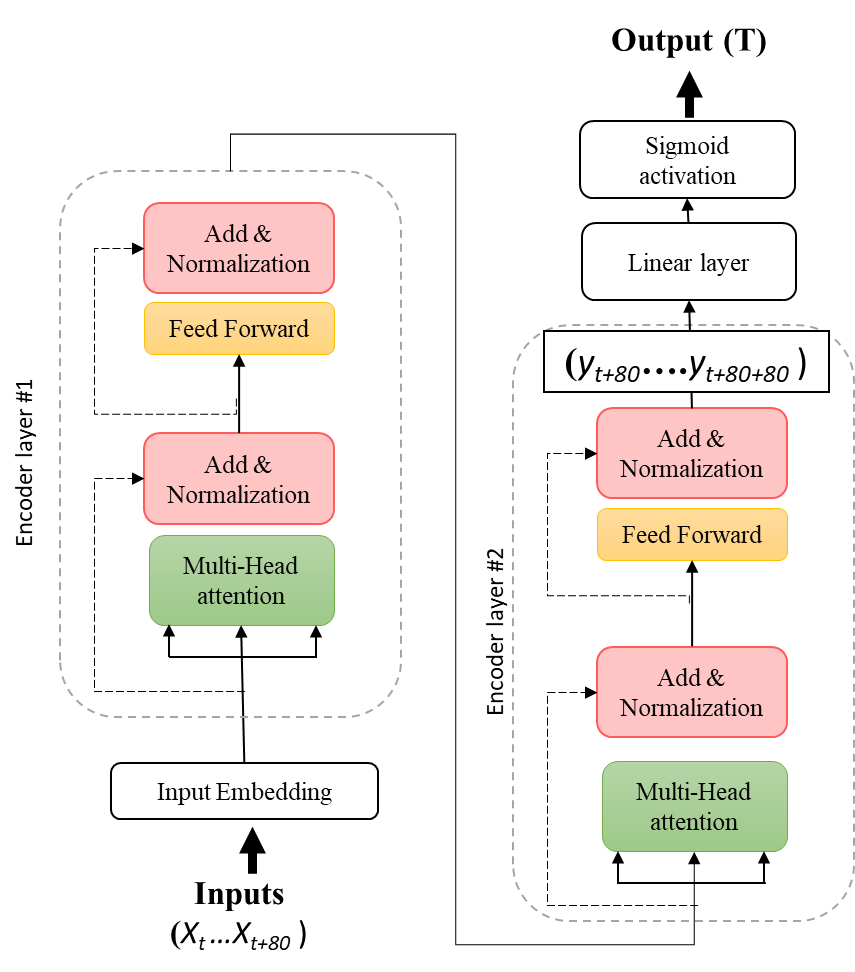}
    \caption{Transformer encoder model architecture \cite{vaswani2017attention} for predicting disruptions with two transformer encoder layers}
    \label{fig:Tr-architecture}
\end{figure}

The transformer encoder model \cite{vaswani2017attention} leverages self-attention mechanisms (in which it allows each token in input sequence to weigh its relationship with all other tokens capturing dependencies regardless of distance) and deep neural networks to learn complex patterns in input data. The architecture of the transformer encoder model, designed for current quench prediction in ADITYA, is illustrated in figure \ref{fig:Tr-architecture}. The transformer encoder architecture incorporates several key components to process and analyse time-series data effectively. The first component, the embedding layer, projects the input sequence of features (in this study six diagnostic signals such as plasma current, soft and hard X-ray outputs, bolometer output, and C111 and H$\alpha$)  into a higher-dimensional space, allowing the model to capture complex relationships within the data. At the core of the architecture are the transformer encoder layers, which consist of two primary sub-components, the self-attention mechanism and a feedforward neural network. The self-attention mechanism evaluates the significance of each feature in relation to others within the sequence by computing attention scores, which aggregate information across the entire sequence and effectively capture contextual dependencies. Following this, the feedforward neural network applies non-linear transformations to the features, enabling the model to learn intricate patterns in the data. To enhance stability during training and mitigate overfitting, layer normalization and dropout regularization are incorporated after each sub-component, contributing to robust training and improved generalization.

The disruption prediction mechanism closely mirrors that of the LSTM model, ensuring consistency in evaluation and enabling reliable performance comparisons between the two architectures. Similar to the LSTM-based approach, the transformer encoder takes an input sequence $X_t$ at time step $t$ and predicts the corresponding label $y_{t+80}$ at time step $t+80$. The final output is generated through a linear layer followed by a sigmoid activation function, producing a single scalar value $T$ for each time step. This value represents the predicted probability of a disruption. To maintain consistency with the LSTM model, a value of 0.5 is applied, predictions where $T > 0.5$ are interpreted as imminent disruption. By adopting the same input-output structure and alarm criterion as the LSTM model, this approach ensures a fair and meaningful comparison between the two architectures. As a result, it offers valuable insights into their respective strengths in predicting disruptions in ADITYA. 
\subsubsection{\label{sec:level1}Training of the models}
Both the models are trained using the binary cross-entropy loss function, which quantifies the difference between the predicted probabilities and the actual labels, where each label indicates whether a disruption occurs at a given time step. This loss function penalizes the model more heavily when the predicted probabilities deviate significantly from the true labels, thereby encouraging the model to generate accurate predictions. To optimize the model parameters during training, the Adam optimizer is employed. To prevent overfitting and improve the model's generalization to unseen data, early stopping is used as a regularization technique. During training, the model's performance is monitored on a validation dataset, and training is halted if the validation loss does not decrease for a specified number of consecutive epochs. This approach ensures that the model does not overfit the training data by learning irrelevant noise, resulting in a model that is both efficient and robust. Together, the binary cross-entropy loss function, Adam optimizer, and early stopping contribute to a stable and well-regularized training process.

For an input sequence $X_t \in \mathbb{R}^{t \times f}$, where $t$ denotes the number of time steps and $f$ is the dimensionality of features at each time step (6 in this case), let $y_t \in \{0, 1\}$ represent the binary ground truth label at time step $t$, and let $\hat{y}_t$ denote the predicted probability at the same time step. The objective is to minimize the binary cross-entropy loss between the predicted output $\hat{y}_t$ and the true label $y_t$. The loss for a single time step is defined as:

\begin{equation}
L_t = -\left[ y_t \log(\hat{y}_t) + (1 - y_t) \log(1 - \hat{y}_t) \right]
\label{eq:bce_single}
\end{equation}

For a sequence of $n$ time steps, the total loss is computed as the average over all time steps:

\begin{equation}
L = \frac{1}{n} \sum_{t=1}^{n} L_t
\label{eq:bce_total}
\end{equation}

The model parameters are updated using the Adam optimizer, which minimizes the loss function by adaptively adjusting the learning rate for each parameter according to the Adam update rule.

\subsubsection{\label{sec:level1}Performance Evaluation Metrics}
\begin{figure*}[!h]
    \centering
    \includegraphics[width=0.8\textwidth, height=12cm]{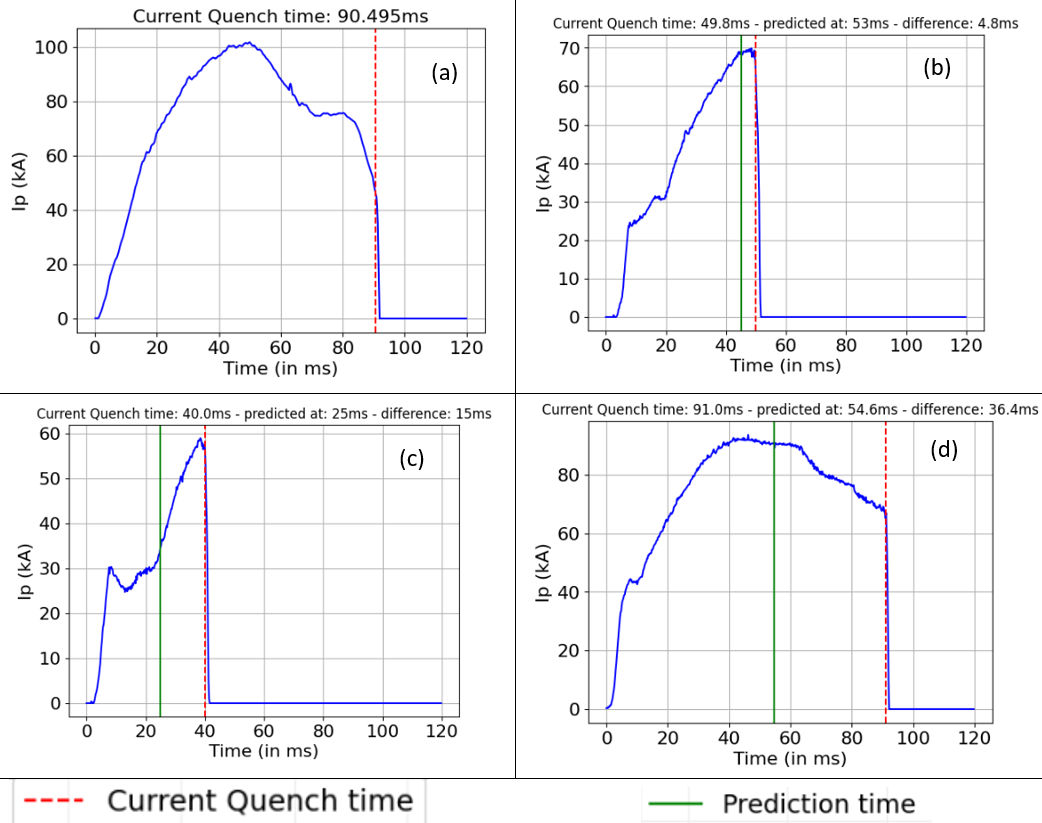}
    \caption{Performance evaluation metrics based on the time window between the prediction time and the prediction threshold, illustrating (a) True negatives, (b) False negatives, (c) True positives, and (d) False positives.}
    \label{Figure:current}
\end{figure*}

To evaluate the performance of the models used in this study, standard evaluation metrics, including precision, recall and accuracy are employed. These metrics provide a quantitative assessment of the model’s ability to make correct predictions, minimize false predictions and accurately classify both disruptive and non-disruptive cases. Performance evaluation is based on the prediction threshold time before the current quench occurs. The quantitative measure of performance comes from the definitions of true positive (TP), true negative (TN), false positive (FP) and false negative (FN) predictions. True positives (TP) refer to disruptive shots that are correctly predicted more than desired threshold value, True negatives (TN) are non-disruptive shots that are not predicted as disruptive and they also include predictions made after 65 milliseconds for non-disruptive shots, as predictions are halted at this point due to current quenching. False positives (FP) represent non-disruptive shots that are incorrectly predicted as disruptive, while false negatives (FN) are disruptive shots predicted less than desired prediction time before the disruption, failing to provide sufficient lead time. Pictorial representation of these metrics for typical plasma current profiles of any tokamak are shown in  figure \ref{Figure:current}.
Precision evaluates the proportion of correctly predicted disruptions (TP)  out of all predicted disruptions (TP + FP), reflecting the model’s ability to minimize false alarms. Recall measures the proportion of actual disruptions (TP) that are successfully predicted, highlighting the model’s ability to detect disruptions effectively. Accuracy assesses the overall correctness of the predictions by calculating the proportion of correctly classified cases (TP + TN) relative to all predictions. These metrics collectively provide a comprehensive evaluation of the model’s performance and its ability to distinguish between disruptive and non-disruptive cases.

\section{\label{sec:level1}Results and Discussion}
This section presents two targeted studies to evaluate and compare the performance of LSTM and Transformer models for disruption prediction in tokamak plasmas using ADITYA data. The first study investigates how model performance varies with different prediction threshold time, by adjusting this threshold from 5 to 20 ms. This investigation is critical for understanding the practical applicability and robustness of machine learning models in timely and accurate disruption prediction. The second study explores the impact of class imbalance by varying the proportion of disruptive to non-disruptive shots in the training dataset, ranging from a naturally balanced distribution to highly biased scenarios.  In real experimental settings, datasets can be biased or non-biased based on the frequency of disruptions, leading to unpredictable class distributions. Studying the effect of such biases will help to evaluate whether ML models can generalize well under realistic conditions, avoid overfitting to dominant class, and maintain reliability in different experimental environments.

\subsection{\label{sec:level1}Based on prediction threshold}
\begin{figure}[h!]
    \centering
    \includegraphics[width=1\linewidth]{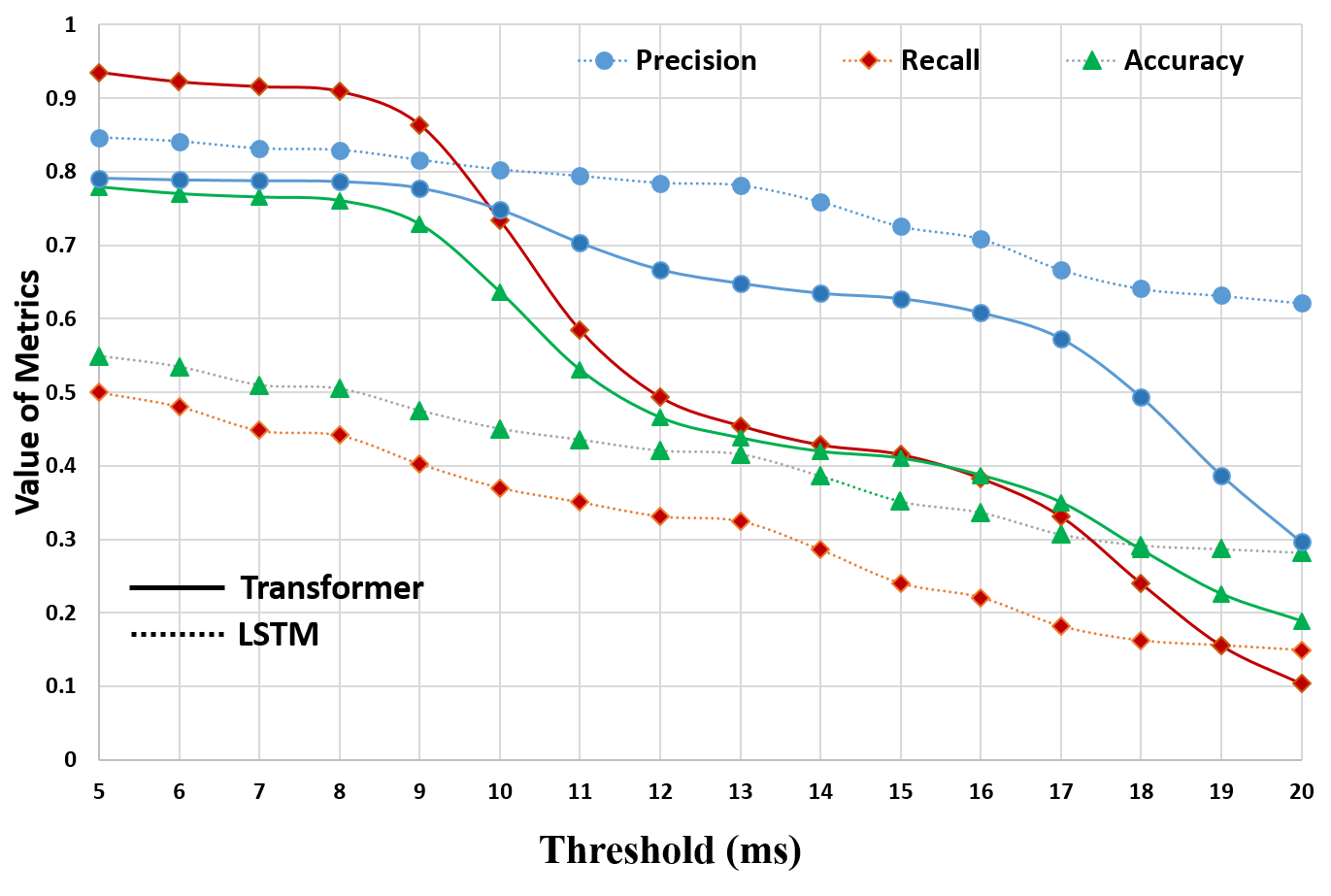}
    \caption{Variation of precision, recall, and accuracy as a function of prediction threshold (ranging from 5 ms to 20 ms) for LSTM and Transformer models}
    \label{fig:result1}
\end{figure}

The data used for this study have original experimental balance for disruptive and non-disruptive experiments and have almost 57$\%$ disruptive and 43$\%$ non-disruptive shots. For this analysis, 30$\%$ of the total 721 shots are allocated for testing. The testing dataset comprised 123 disruptive shots and 93 non-disruptive shots. To evaluate the performance of both models, the prediction threshold (the time window or warning time for disruption prediction) is varied from 5 ms to 20 ms. figure \ref{fig:result1} illustrates the performance metrics (precision, recall and accuracy) of the transformer and LSTM models as the prediction threshold increases from 5 ms to 20 ms. While the transformer outperforms LSTM in recall and accuracy consistently across all threshold values, its precision is slightly lower. Although both models show reduced recall and accuracy with increasing thresholds, the transformer maintains greater stability and exhibits a slower decline in performance than the LSTM, particularly up to the 8-10 ms prediction threshold.
\begin{figure*}[!h]
    \centering
    \includegraphics[width=0.8\textwidth, height=12cm]{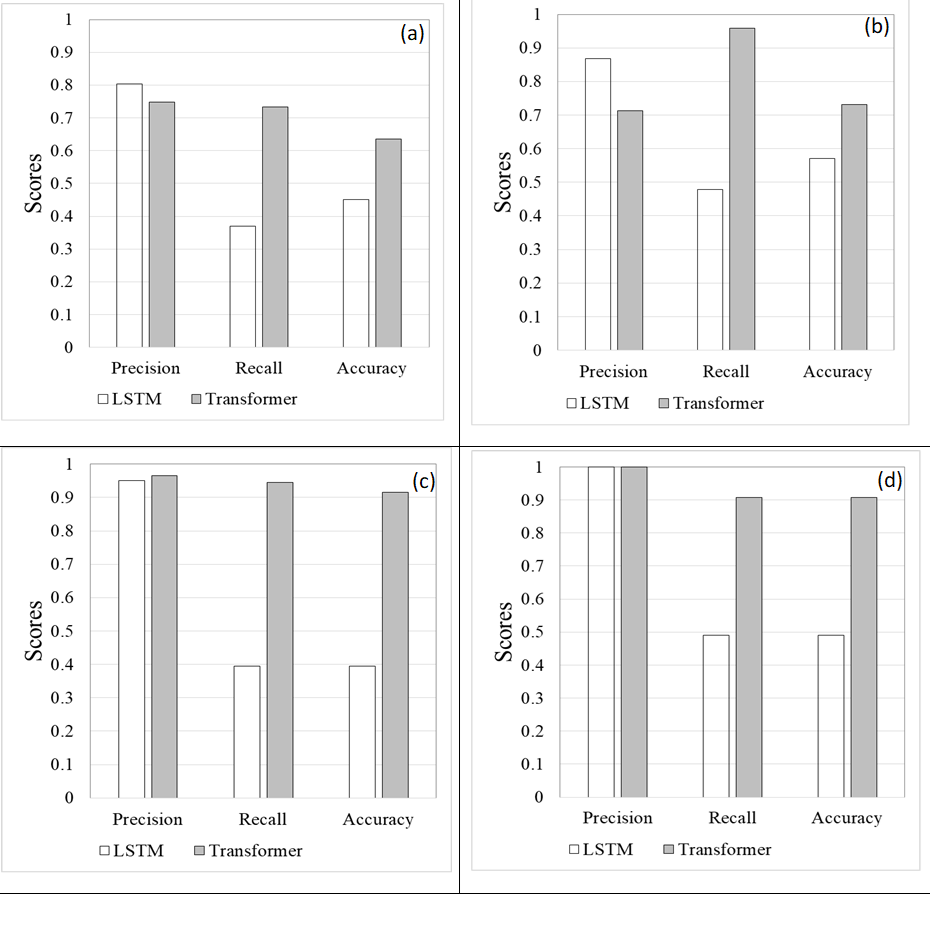}
    \caption{Performance comparison between Transformer and LSTM models across different ratios of disruptive and non-disruptive shots - (a) natural data (unbiased), (b) mildly biased (60$\%$ disruptive), (c) biased (70$\%$ disruptive), and (d) highly biased (80$\%$ disruptive)}
    \label{Figure:result2}
\end{figure*}
Overall, the transformer model proves to be more robust and reliable for disruption prediction. It's superior recall and accuracy, along with a more gradual decline in precision, make it a preferable choice over the LSTM. However, both models highlight a trade-off between precision and recall as the prediction window widens, underscoring the need to carefully balance prediction threshold selection based on specific application requirements.
This study also suggests that the recall value is significantly higher for the transformer model compared to the LSTM, making the transformer a reliable alternative for tokamak scientists. A high recall indicates the model's ability to minimize instances where disruptions occur but go undetected. For large machines like ITER, any disruption event can be extremely costly due to the potential damage caused by the release of high energy during such events, but this is also significant for small tokamak like ADITYA, as any undetected disruption event causes damage in limiters and subsequently harm the successive experiments by increasing the impurities in vacuum vessel. Therefore, a model with high recall is highly desirable and the performance of transformer  demonstrates it's capability to meet this critical requirement for ADITYA tokamak data, while it is required to confirm the same with data from long duration tokamaks. 

\subsection{\label{sec:level1}Performance evaluation based on the data biasing for disruptive shots}
Based on the performance comparison of both models presented above, we proceed with a prediction threshold of 8 ms. This threshold is deemed appropriate for mitigating disruptions in the ADITYA tokamak, considering its maximum plasma duration of 120 ms. In this section, we analyse the impact of varying the proportion of disruptive and non-disruptive training shots on model performance. In addition to the natural ratio of disruptive to non-disruptive shots, three more scenarios are considered, 60$\%$ disruptive shots (mildly biased), 70$\%$ disruptive shots (biased), and 80$\%$ disruptive shots (highly biased). To ensure consistency, a dataset of 510 shots is used for this analysis, with 70$\%$ (360 shots) allocated for training and 30$\%$ (150 shots) for testing in each scenario. It is important to note that this dataset is smaller than the 721 shots used in the earlier comparison of LSTM and transformer encoder models. This choice is made to maintain an adequate number of non-disruptive shots in each scenario while keeping the total dataset size constant, ensuring fairness and uniformity in the evaluation.

Figure \ref{Figure:result2} compares the performance of the transformer and LSTM models across different ratios of disruptive and non-disruptive shots - (a) natural data, (b) mildly biased (60$\%$ disruptive shots), (c) biased (70$\%$ disruptive shots), and (d) highly biased (80$\%$ disruptive shots). The transformer consistently outperforms the LSTM in all scenarios, demonstrating greater robustness to variations in data distribution. In the natural data scenario, the transformer exhibits higher recall, and accuracy, with a particularly significant advantage in recall, indicating better reliability in detecting disruptions. As the dataset becomes increasingly biased toward disruptive shots, both models show an improvement in precision, reflecting enhanced accuracy in identifying disruptions. The transformer's performance remains more balanced across precision, recall, and accuracy. Such consistency suggests that the transformer model is less sensitive to changes in the proportion of disruptive and non-disruptive shots, making it reliable for real-world applications where data distributions can be imbalanced or variable. This robustness is particularly critical for disruption prediction in tokamak experiments, where biased datasets may arise naturally due to operational conditions or experimental setups. This demonstrates its adaptability and reliability, making it a more suitable choice for disruption prediction in tokamak experiments. Also, in the context of plasma disruption prediction, a higher recall value indicates fewer false alarms, which is desirable to avoid unnecessary shutdowns of the experiment. The trends also suggest that for the transformer model, the F1 score improves as the dataset becomes more biased. 

\section{\label{sec:level1}Conclusion and Future Work}

Our work demonstrates the effectiveness of transformer-based deep learning models for early prediction of current quench events in the ADITYA tokamak, a critical step toward proactive disruption mitigation. Compared to LSTM, the transformer model consistently delivers higher recall, improved accuracy, and greater robustness across varying prediction thresholds and data biases. Importantly, this work also presents a quantitative comparison between transformer and LSTM architectures, offering insights into their relative strengths and limitations. A key observation is that the transformer model maintains stable and high performance even as the current quench prediction threshold time increases, whereas the LSTM model shows a consistent decline in accuracy and recall under the same conditions. This robustness to increasing lead times highlights the superior ability of transformers to capture long-range dependencies in multivariate plasma signals, an essential trait for early disruption forecasting. Additionally, the test dataset used in this study is carefully balanced, with comparable numbers of disruptive and non-disruptive shots. This ensures that the model evaluation is not artificially inflated due to class imbalance. As demonstrated in our analysis, biasing the dataset toward more disruptive cases can deceptively boost model performance, especially in recall and accuracy. By maintaining a balanced test distribution, we ensure a more rigorous and realistic assessment of generalization performance.
To the best of our knowledge, this is the first application of transformer models to ADITYA tokamak data, highlighting their potential for short-pulse devices. In future work, we aim to (i) assess real-time inference capabilities of the transformer model, (ii) evaluate its applicability to longer-duration tokamaks, and (iii) incorporate attention-based interpretability to better understand the role of specific diagnostics. Expanding the dataset and benchmarking against additional baselines will further strengthen generalization of the model and reliability across diverse operational scenarios. The demonstrated superiority of transformers, especially in terms of recall, suggests a viable pathway toward real-time disruption prediction tools, potentially applicable to diverse tokamak configurations. These directions will further enhance the reliability and utility of AI-assisted disruption mitigation tools in fusion plasma research.
- Although current quench appears as a sudden and localized phenomenon, its precursors are embedded in the plasma's earlier states. This finding supports the feasibility of early current quench prediction, emphasizing the potential for proactive correction rather than mere mitigation. Such a model can also be expected to enhance its performance in long-duration plasma machines like ITER, DEMO etc. without significantly affecting the overall outcome.

\section*{Acknowledgement}
We acknowledge the support of the ADITYA tokamak team, IPR to provide necessary data for this research work.

\bibliographystyle{elsarticle-num}
\bibliography{aipsamp.bib}

\end{document}